\documentclass{article}

\usepackage[labelfont=bf]{caption} % Bold figure number
\usepackage[font={small}]{caption}   %% FIGURE CAPTION FONT MODIFICATION

\usepackage{graphicx}% Include figure files
\usepackage{dcolumn}% Align table columns on decimal point
\usepackage{bm,bbm}% bold math
\usepackage[pdftex,colorlinks=true,linkcolor=blue,citecolor=blue]{hyperref}
\usepackage[utf8]{inputenc}
\usepackage{lipsum}  
\usepackage{chngpage}
 \usepackage{amsmath}
\usepackage{amssymb}
\usepackage{authblk}
\usepackage{amsfonts}
\usepackage{placeins}
\usepackage[numbers, sort&compress, super]{natbib}  %%% For compressing the reference numbering

\newcommand{\onlinecite}[1]{\hspace{-1 ex} \nocite{#1}\citenum{#1}}
    \usepackage{lineno}   %% <- no mathlines option
    \usepackage{amsmath}  %% <- after lineno
    \usepackage{etoolbox} %% <- for \cspreto, \csappto

    \newcommand*\linenomathpatch[1]{%
    \cspreto{#1}{\linenomath}%
    \cspreto{#1*}{\linenomath}%
    \csappto{end#1}{\endlinenomath}%
    \csappto{end#1*}{\endlinenomath}%
    }

    \linenomathpatch{equation}
    \linenomathpatch{gather}
    \linenomathpatch{multline}
    \linenomathpatch{align}
    \linenomathpatch{alignat}
    \linenomathpatch{flalign}
\begin{document}

\title{ Evidence for Higher order topology in Bi and Bi$_{0.92}$Sb$_{0.08}$}

\author[1]{Leena Aggarwal}
\author[2]{Penghao Zhu}
\author[2]{Taylor L. Hughes}
\author[1*]{Vidya Madhavan}

\affil[1]{\footnotesize{Department of Physics and Materials Research Laboratory, University of Illinois
Urbana-Champaign, Urbana, Illinois 61801, USA}}
            %, University of Illinois at Urbana-Champaign, Urbana, IL, USA}}
\affil[2]{\footnotesize{Department of Physics and Institute for Condensed Matter Theory, University of Illinois at Urbana-Champaign, Urbana, Illinois 61801, USA}}
            %, University of Illinois at Urbana-Champaign, Urbana, IL, USA}}
\affil[*]{\footnotesize{To whom correspondence should be addressed: vm1@illinois.edu}}

%\date{\today}

\maketitle

\begin{abstract}
Higher order topological insulators (HOTIs) are a new class of topological materials which host protected states at the corners or hinges of a crystal.  HOTIs provide an intriguing alternative platform for helical and chiral edge states and Majorana modes, but there are very few known materials in this class. Recent studies have proposed Bi as a potential HOTI, however, its topological classification is not yet well accepted. In this work, we show that the (110) facets of Bi and BiSb alloys can be used to unequivocally establish the topology of these systems. Bi and Bi$_{0.92}$Sb$_{0.08}$ (110) films were grown on silicon substrates using molecular beam epitaxy and studied by scanning tunneling spectroscopy. The surfaces manifest rectangular islands which show localized hinge states on three out of the four edges, consistent with the theory for the HOTI phase. This establishes Bi and Bi$_{0.92}$Sb$_{0.08}$ as HOTIs, and raises questions about the topological classification of the full family of Bi$_{x}$Sb$_{1-x}$ alloys.
\end{abstract}

\newpage
\vspace{12pt}
\section*{Introduction}
\quad \ HOTIs are topological crystalline insulators that have protected  topological  features on the boundaries with co-dimension greater than one. In contrast to (1st order) three-dimensional time-reversal protected topological insulators that show an insulating bulk with conducting surface modes, three-dimensional 2nd order HOTIs show insulating bulk and surfaces, but with one-dimensional conducting channels at hinges (i.e., the intersection of two surface facets) or equivalent crystal defects  \cite{schindler2018higher,song2017d,benalcazar2018prb,Schindlereaat0346,Olsen2020gaplss,zhu2020quantized}.  While several groups have recently succeeded in observing HOTI phases in  electronic circuit, phononic, and photonic systems \cite{serra2018observation,peterson2018quantized,noh2018topological,imhof2018topolectrical}, the experimental realization and characterization of HOTIs in solid state materials has proven more challenging.  Among the  proposed HOTI materials \cite{Schindlereaat0346,choi2020evidence,lee2020two,zhang2020mobius,wang2019higher,zhang2019higher,park2019higher}, bulk Bi has recently emerged as a key candidate. Bi and Bi$_{1-x}$Sb$_x$ alloys are well-known important topological materials with highly tunable electronic properties.  Scanning tunneling microscopy studies \cite{schindler2018higher,drozdov2014one} have observed signatures of one-dimensional modes on the edges of (111) facets of Bi, but the higher order topological origin of these edge modes remains to be unequivocally established. Indeed an earlier study modeled the (111) facet as a free-standing Bi-bilayer \cite{drozdov2014one}, which is predicted to be a quantum spin Hall insulator, and hence attributed the observed hinge modes to edge states of a quantum spin Hall system. In contrast, a recent study \cite{schindler2018higher} ascribed the metallic edge modes to the manifestation of a bulk HOTI protected by crystalline and time-reversal symmetries.  New experimental data is necessary to distinguish between these scenarios. Importantly, the confirmation of Bi as a HOTI  would indicate that the low-energy bands of bismuth must have trivial time-reversal protected $\mathbbm{Z}_{2}$ topology \cite{schindler2018higher,hsu2019topology}. This not only has implications for the topology of bulk Bi, which still remains controversial, but also the family of Bi$_{1-x}$Sb$_{x}$ alloys.
%\begin{figure}[h!] 

%	\centering
	%	\includegraphics[width=0.7\columnwidth]{f-1r.pdf}
    %	\caption{\textbf{Atomic structure and STM measurements on Bi(110) and Bi$_{0.92}$Sb$_{0.08}$(110) films grown on silicon substrates:} (a) Crystal structure of Bismuth with primitive cell, showing rhombohedral vectors  ($\vec{a}_{1}$,$\vec{a}_{2}$,$\vec{a}_{3}$). The (110) plane is indicated by a yellow rectangular box. (b,c) Large area (500 nm $\times$500 nm) topographic images of Bi(110) and Bi$_{0.9}$Sb$_{0.1}$(110) films respectively. (d) Top view of Bi(110) plane (as marked yellow rectangular box in Figure 1(a)) with lattice constants. (e,f)  STM images of 10 nm $\times$10 nm  area on the rectangular facets of Bi(110) and Bi$_{0.9}$Sb$_{0.1}$(110) films (at 150 pA, 300 mV) respectively. (g,h) Height (`h') profiles along the line cuts (marked in red and blue) show the lattice periodicity of the pseudo-cubic crystal structure of Bi(110). (i,j) Representative $dI/dV$ spectra near the centers of Bi(110) (ac modulation 3.5 mV and current 80 pA) and Bi$_{0.9}$Sb$_{0.1}$(110) (ac modulation 3.0 mV and current 120 pA) islands respectively. The spectra are vertically offset for clarity.}
%	\label{f1}
%\end{figure}
\section*{Results}
\textbf{STM measurements on Bi and Bi$_{0.92}$Sb$_{0.08}$ films.}
Bi naturally cleaves along the (111) plane, which makes STM studies of other planes, e.g, (110), difficult in bulk cleaved crystals.  To circumvent this problem, we use molecular beam epitaxy to grow Bi and Bi$_{0.92}$Sb$_{0.08}$ films on n-doped silicon (111) wafers  \cite{takayama2015one,kawakami2015one,nagao2004nanofilm,PhysRevB.91.075429,2020APExp..13h5506N}. The films were grown in a custom MBE system (described in more detail in the Methods section), and the composition was confirmed using Rutherford Backscattering Spectrometry (RBS) (see Supplementary Note 4). Film thicknesses were chosen to be in the range of 30-90 BLs to ensure that they are representative of the bulk. 

To establish a benchmark for comparison, we first reproduce the edge state data previously obtained on Bi(111) crystals using our films. STM topographies on our Bi(111)/Si(111) films reveal triangular islands and a hexagonal lattice indicative of (111) facets. The dI/dV spectra on the bulk and edges of the triangular islands show signatures of the Van Hove singularity associated with the surface states, as well as the emergence of clear 1D modes at type-A edges (see Supplementary Note 1), as was observed in the cleaved bulk Bi crystals \cite{drozdov2014one}. Having confirmed the presence of edge modes in the (111) films, we move on to the (110) films.

\begin{figure}[h!]
	\centering
		\includegraphics[width=0.65\columnwidth]{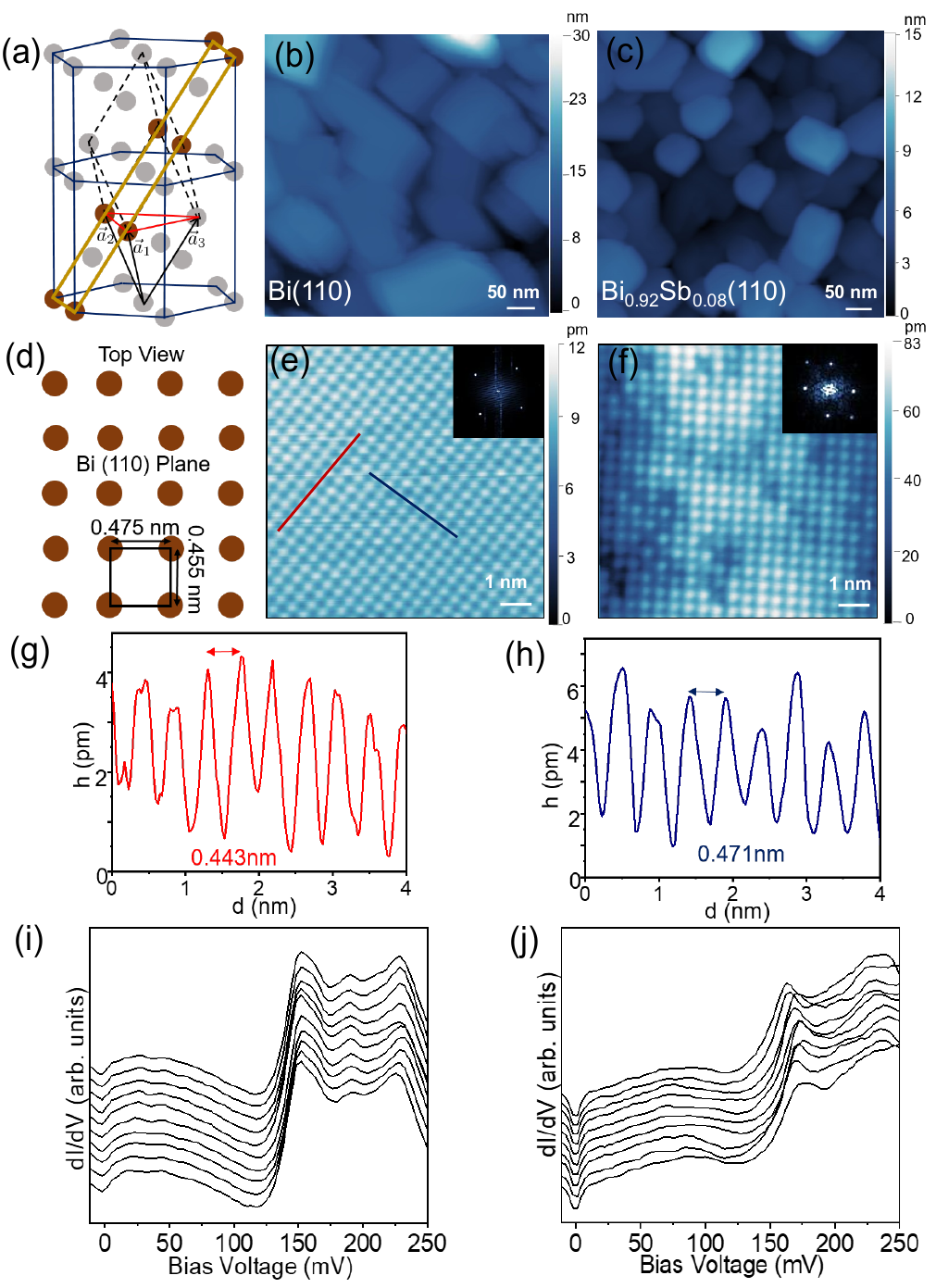}
    	\caption{\textbf{Atomic structure and STM measurements on Bi(110) and Bi$_{0.92}$Sb$_{0.08}$(110) films grown on silicon substrates:} (a) Crystal structure of Bismuth with primitive cell, showing rhombohedral vectors  ($\vec{a}_{1}$,$\vec{a}_{2}$,$\vec{a}_{3}$). The (110) plane is indicated by a yellow rectangular box. (b,c) Large area (500 nm $\times$500 nm) topographic images of Bi(110) and Bi$_{0.92}$Sb$_{0.08}$(110) films respectively. The color bars show the relative height of the topography. (d) Top view of Bi(110) plane (as marked yellow rectangular box in Figure 1(a)) with lattice constants. (e,f)  STM images of 9 nm $\times$9 nm  area on the rectangular facets of Bi(110) and Bi$_{0.92}$Sb$_{0.08}$(110) films (at 150 pA, 300 mV) respectively, with color bars indicating the relative height. Insets show the Fast Fourier Transform (FFT) Topography (g,h) Height (`h') profiles along the line cuts (marked in red and blue) show the lattice periodicity of the pseudo-cubic crystal structure of Bi(110). (i,j) Representative $dI/dV$ spectra near the centers of Bi(110) (ac modulation 3.5 mV and current 80 pA) and Bi$_{0.92}$Sb$_{0.08}$(110) (ac modulation 3.0 mV and current 120 pA) islands respectively. The spectra are vertically offset for clarity.}
	\label{f1}
\end{figure}

By changing the growth conditions (see Methods section), we were able to grow Bi and Bi$_{0.92}$Sb$_{0.08}$ in the (110) orientation on the same Si(111) substrates. STM images of the (110) oriented films of both Bi and Bi$_{0.92}$Sb$_{0.08}$ are shown in Figure \ref{f1}. Large area topographic images of bismuth films (Figure \ref{f1} (b)) show that the surface is uniformly covered with (110) oriented rectangular islands with longer edges parallel to the ($1\bar{1}0$) direction. Figure \ref{f1} (e) shows an STM image with its corresponding Fourier transform in the inset. Height profiles along line cuts (red and blue color lines in Figure \ref{f1} (e)) confirm the lattice constants of the pseudo-cubic crystal structure of Bi(110), i.e., 0.443 nm and 0.471 nm in Figure \ref{f1} (g) and (h), respectively. Representative spectra at the center of the Bi and Bi$_{0.92}$Sb$_{0.08}$ (110) islands, shown in Figure \ref{f1} (i) and (j), reveal a broad increase in the density of states (DOS) above 130 mV in both samples. This increase can be ascribed to the spin-split Rashba surface states known to exist in bismuth and its alloys \cite{takayama2015one,takayama2014rashba}. 
\\

\begin{figure}[h!]
	\centering
		\includegraphics[width=0.75\textwidth]{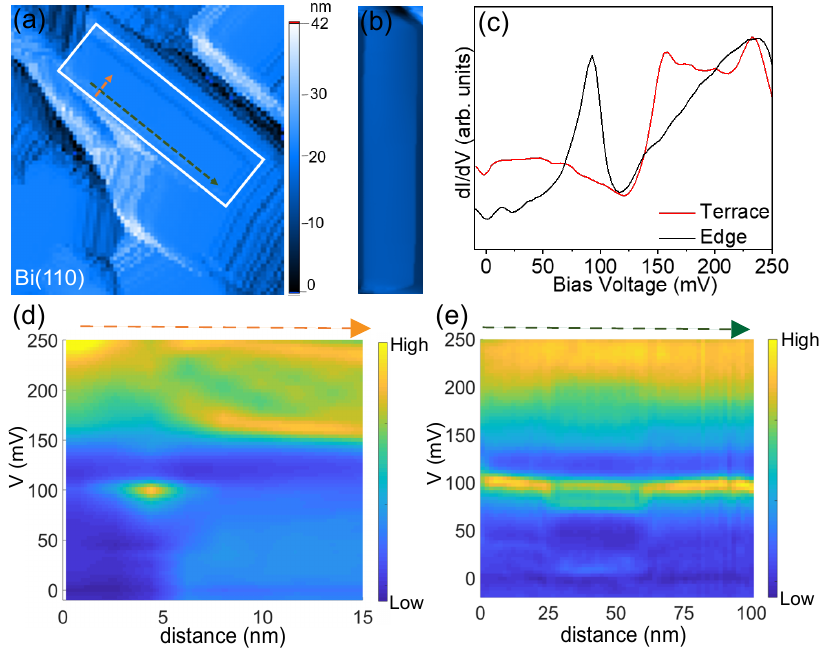}
\caption{\textbf{Spectroscopic measurements along and across the edge of a rectangular facet of a Bi(110) island:} (a) Large area (250 nm $\times$350 nm) image of a Bi(110) film. The island of interest is marked in white. (b)  Zoomed in image of the Bi(110) island in Figure 2 (a). The color bar shows the relative height in the topography. (c) dI/dV spectra at the edge and center of the island  (ac modulation 3.5mV and current 100 pA). The edge spectrum shows a peak corresponding to the edge mode. (d,e) Intensity plot of STM spectra across (orange dashed line in figure 2(a)) and along (green dashed line in figure 2(a)) the step edge (ac modulation 3.5mV and current 100 pA). The color bars indicate the relative height of the DOS. As the linecut in 2 (d) shows, the peak at ~100 meV which corresponds to the edge mode, is clearly localized to within a few nms of the edge.}
	\label{f2}
\end{figure}

\begin{figure*}[t]
	\centering
		\includegraphics[width=1.0\textwidth]{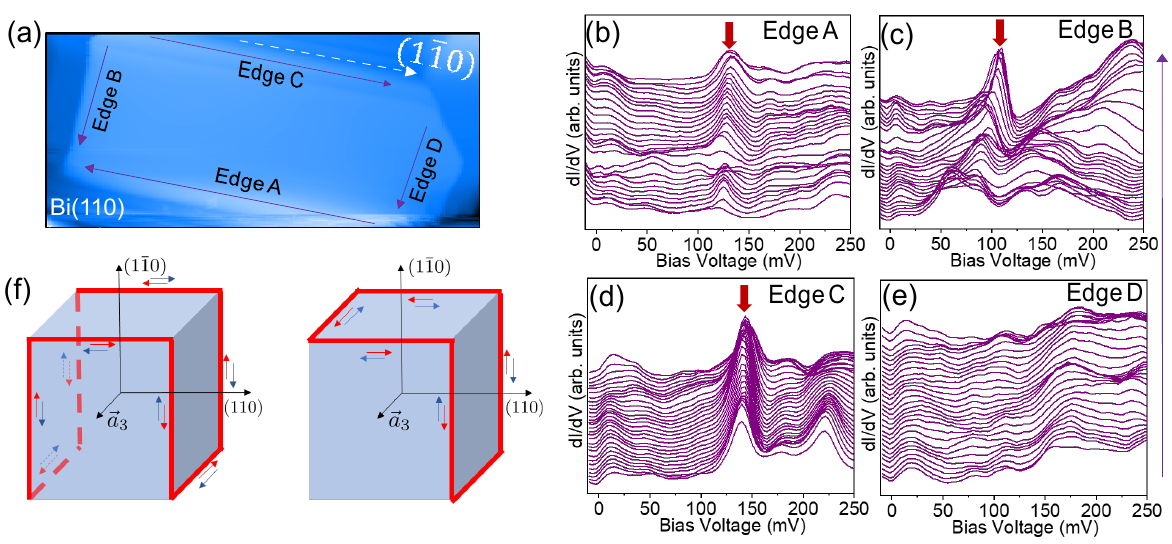}
\caption{\textbf{Metallic modes at the edges of rectangular island of Bi(110):} (a) Topographic image of a Bi(110) rectangular island of 168 nm $\times$ 80 nm area. The four edges of the island are marked as Edge A, Edge B, Edge C, and Edge D, respectively. The dotted white arrow shows the (1$\bar{1}$0) direction. (b,c,d,e) STM spectra (at ac modulation 3.5 mV and current 60 pA) along the four edges (as shown by purple arrows in Figure 3 (a)). A constant slope was subtracted from each spectrum for clarity. The raw data without slope subtraction is shown in supplementary Figure 3 (b,c,d,e). The red block arrows indicate the sharp peaks corresponding to the edge modes that are only seen on three of the four edges.  (f) Two possible configurations of helical hinge modes for Bi in the HOTI phase, where three edges of a rectangular island on a (110) facet hold helical modes. The $\vec{a}_{3}$ indicates a lattice vector of the Bi crystal as shown in Figure \ref{f1} (a).}
	\label{f3}
\end{figure*}

\noindent\textbf{Spectroscopic measurements on a rectangular facet of Bi(110).}
We first focus on the edge states observed on rectangular islands of the Bi (110) films. Spectra on the edge of an island show a sharp peak in the DOS (density of states)  as shown in Figure \ref{f2} (c). This singularity corresponds to bound modes which are localized within a few nanometers of the edge, as seen in the spectroscopic measurements across and along the edge shown in Figure \ref{f2} (d) and Figure \ref{f2} (e) respectively. This sharp peak resembles the DOS feature observed on the edges of Bi(111). To further characterize this edge mode, we obtain spectroscopic data on all four edges of a (110) oriented rectangular island as shown in Figure \ref{f3} (b,c,d,e). Intriguingly we find that only three out of four edges show clear edge modes, while the fourth edge shows no peaks. We confirmed this observation by measurements on different rectangular facet islands on multiple samples. In all cases, the edge that shows no edge modes is  the one that is perpendicular to the ($1\bar{1}0$) direction. Crucially, the observation of edge states on three out of the four edges cannot be attributed to different bonding on just one side, which was a key part of the argument in Ref.~\onlinecite{drozdov2014one}.  By looking at the lattice structure of the ($110$) facet, we know that the two short sides are identical from a bonding perspective (see Supplementary Note 6). Yet, for a given island, one short side shows hinge states while the other does not.
\\

\noindent\textbf{Theoretical Justification.}
To understand the novel pattern of rectangular island edge modes on the (110) facets, let us first briefly review the recent developments for the topology of bismuth. Specifically, Refs.~\onlinecite{schindler2018higher} and \onlinecite{hsu2019topology} argued that the bulk topology of bismuth can be described as two copies of a strong 3D TI, i.e., two band inversions at the $T$-point. 
  Since there are an even number of strong TI copies, they conclude that the low-energy bulk bands of bismuth have trivial strong $\mathbbm{Z}_{2}$ topology,  but may support two robust Dirac points on high-symmetry surface facets exhibiting additional crystalline symmetries. For example, the (111) facet has $\hat{C}_{3}$ rotation symmetry; the two Dirac points are located at the $T$-point and belong to different $\hat{C}_{3}$ eigenvalue sectors having $\hat{C}_{3}$ eigenvalues $-1$ and $\exp(\pm i\pi/3)$ respectively \cite{schindler2018higher}. If the two Dirac points are pinned at the same energy, then they cannot be gapped out while preserving the $\hat{C}_{3}$ (no inter-cone gap) and  time-reversal symmetries (no intra-cone gap). However, there are no symmetry constraints that fix the energies of the two cones, and generically it is expected that the (111) surface will be gapped \cite{schindler2018higher}. In this context, Ref. \onlinecite{schindler2018higher} considered the case where the (111) facet is the top facet of a hexagonal crystal (as shown in Figure \ref{f1} (a)). For this geometry the $\hat{C}_{3}$ symmetry and time-reversal symmetry are compatible with sign-alternating surface-state masses/gaps on the side facets (this breaks $\hat{C}_{6}$ but preserves $\hat{C}_{3}$) \cite{schindler2018higher}. Since the surface Dirac cones on the top/bottom (111) facets can also be generically gapped while preserving the symmetries, this will lead to sharply localized hinge modes where the side surfaces intersect the top/bottom surfaces, which were first observed for bulk Bi crystals in Ref. \onlinecite{drozdov2014one,schindler2018higher}, and confirmed for Bi films in our measurements above.
  
We propose that the observed edge states of an island on the (110) surface can be ascribed to  $\hat{C}_{2}$ (of the bulk crystal around the $(1\bar{1}0)$ axis) and time-reversal symmetry protected higher order topology consistent with Ref. \onlinecite{hsu2019topology}. The (110) surfaces of our islands do not have any protective crystal symmetry and thus we generically expect the two surface Dirac cones on those surfaces (as predicted by Ref. \onlinecite{hsu2019topology})  to be gapped. When considering hinge states though, we must also consider the mass gaps for surface states on the side-surfaces of such islands. First, the two long sides/edges of the (110) rectangular islands do not have any protected crystal symmetry, hence we expect the surface states to be gapped. Second, the two short sides/edges of the (110) rectangular islands are truncated ($1\bar{1}0)$ surfaces, which could show remnants of the two gapless Dirac cone surface states.  Indeed, Ref. \onlinecite{hsu2019topology} recently emphasized that the ($1\bar{1}0$) surface has $\hat{C}_{2}$ rotation symmetry that protects the two, generically unpinned, surface Dirac points related by $\hat{C}_2$ rotation. In this case, the ($1\bar{1}0$) facet, cannot be gapped out (perturbatively) as long as the $\hat{C}_{2}$ and time-reversal symmetries are preserved since the Dirac cones do not coincide at a TRIM. However, since we are considering a surface island, translation symmetry is strongly broken on the ($1\bar{1}0$) facets. Hence, we do not expect the surface states to be robust since their momentum space locations become irrelevant. 
From these arguments we expect that the surfaces that form the top and sides of the island can be generically gapped, so we can explore our observations in the context of higher order topology in this system.  In contrast to previous systems with crystalline-protected topology, (e.g., the mirror-protected surface Dirac cones in SnTe \cite{tanaka2012experimental}), the $(110)$ surface of Bi does not preserve the $\hat{C}_{2}$ bulk symmetry responsible for the higher order protection\cite{Schindlereaat0346}. However, this does not mean that we cannot observe signatures of higher order topology, it just implies that the possible hinge state patterns may be affected by the details of the surface termination. For a rectangular crystal as shown in Figure \ref{f3} (f), where ($1\bar{1}0$) is the top edge/facet, there are two inequivalent configurations of surface mass gaps that match our observations of hinge modes on the $(110)$ facet; i.e., that helical modes are localized on only three edges out of four. The configuration of helical hinge modes shown in the left panel of Figure \ref{f3} (f) nominally preserves both $\hat{C}_{2}$ (on the ($1\bar{1}0$) surface termination) and time-reversal symmetries, while the configuration shown in the right panel of Figure \ref{f3} (f) preserves only the time-reversal symmetry (details can be found in Supplementary Note 2). In addition, both configurations of helical hinge modes in Figure \ref{f3} (f) that are consistent with our observations require broken inversion symmetry on the surface termination (which is a symmetry of the bulk of a bismuth crystal). However, this is not problematic since inversion is naturally broken for an island on a surface. Thus, we find that our observations corroborate, and are clearly indicative of, higher order topology  in bismuth.
\\

\begin{figure*}[t]
	\centering
		\includegraphics[width=1.0\textwidth]{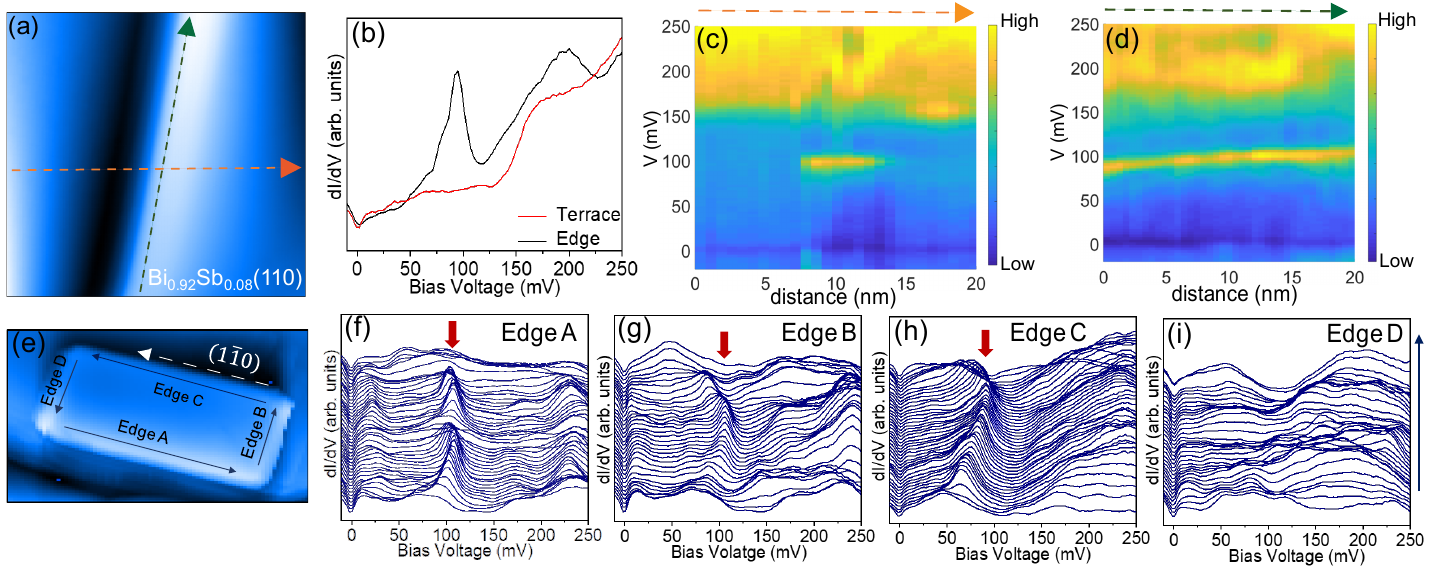}
	\caption{\textbf{STM data on Bi$_{0.92}$Sb$_{0.08}$(110) film:} (a) STM topographic image of 20 nm $\times$ 20 nm area of Bi$_{0.92}$Sb$_{0.08}$(110) film, showing an island edge. (b) dI/dV spectra at the edge and over the island terrace (at ac modulation 3.5mV and current 100 pA), showing the peak around 100 meV corresponding to the mode at the edge . (c,d) dI/dV spectra obtained across (orange dashed line in Figure 4(a)) and along (green dashed line in figure 2(a)) the island edge, respectively. The color bars indicate the relative height of density of states. (e)  Topography of rectangular facet of Bi$_{0.92}$Sb$_{0.08}$(110) with 50 nm $\times$ 25 nm area. The four edges are marked as Edge A, Edge B, Edge C, and Edge D, respectively. The white arrow along the edge shows the (1$\bar{1}$0) direction. (f,g,h,i)  Spectroscopic data (at ac modulation 3 mV and current 120 pA) along the  four edges A, B, C,D respectively (following the blue arrows in Figure 4 (e)). A line with constant slope was subtracted from all spectra for clarity. The raw data for these figures without slope subtraction is shown in supplementary Figure 3 (g,h,i,j). The red block arrows indicate the sharp peaks corresponding to the edge modes that are only seen on three of the four edges. }
	\label{f4r}
\end{figure*}

\noindent \textbf{Spectroscopic measurements on Bi$_{0.92}$Sb$_{0.08}$(110).}
We can now use our established signature of higher order topology in Bi to determine the topological nature of Bi alloyed with Sb. Previous theoretical studies have suggested that the Bi$_{1-x}$Sb$_{x}$ alloy undergoes a trivial-to-topological band-inversion transition at approximately $x= 0.04$ \cite{lenoir1996bi,fu2007topological,teo2008surface}. However, similar to Bi, the narrow band gaps and the existence of multiple bulk and surface bands near the Fermi energy have made the theoretical and experimental determination of the topological properties of Bi$_{1-x}$Sb$_{x}$ complicated. From our studies on Bi, we propose that the pattern of the edge modes on the (110) oriented rectangular islands may provide a clear picture of the topology in this compound. Figure \ref{f4r} (b) shows STM spectra at an edge and plateau of the (110) facet of Bi$_{0.92}$Sb$_{0.08}$. A peak centered around 100meV is observed at the edge, similar to that observed on Bi islands. Spectroscopic measurements across and along the edge shown in Figure \ref{f4r} (c) and Figure \ref{f4r} (d) respectively, reveal that this mode is localized to the edge. Importantly, similar to our data on Bi, we observe these edge modes on three out of the four edges of (110) oriented rectangular islands as shown in Figure \ref{f4r} (f,g,h,i). The explanation of this pattern of edge modes requires the system to be higher order, and hence requires that the bulk has trivial strong $\mathbbm{Z}_{2}$ topology. Hence, our data on Bi$_{0.92}$Sb$_{0.08}$ suggests that it has trivial strong $\mathbbm{Z}_{2}$ topology and falls in the HOTI regime. This bears further study using other experimental techniques since it calls into question some long-accepted predictions about BiSb alloys, and indeed may have implications for the celebrated identification of nearby Bi$_{0.9}$Sb$_{0.1}$ as a strong $\mathbbm{Z}_{2}$ topological insulator. 

\section*{Discussion}
\quad \ In summary, we have studied (110) oriented Bi and Bi$_{0.92}$Sb$_{0.08}$ films with STM and STS. We observe bound modes on three out of four edges of a rectangular island on (110) facets. Moreover, the edge showing no edge modes is always perpendicular to the ($1\bar{1}0$) direction. Theoretically, the novel pattern of the metallic edge modes observed on the (110) facets can be explained by a higher order topology protected by time reversal ($\hat{T}$) symmetry and bulk two-fold rotation ($\hat{C}_{2}$) symmetry around the ($1\bar{1}0$)-axis \cite{hsu2019topology}. Based on this we confirm that Bi and Bi$_{0.92}$Sb$_{0.08}$ belong to the HOTI class of topological materials. The latter result is particularly interesting since previous theoretical studies indicate that Bi$_{1-x}$Sb$_{x}$ with $x \sim 0.08$ should already be in a strong $\mathbbm{Z}_{2}$ topological insulator phase heralded by a band inversion predicted at $x\sim 0.04$ \cite{lenoir1996bi,fu2007topological,teo2008surface}. Our work brings new insights to the topology of Bi$_{0.92}$Sb$_{0.08}$ which is revealed as a new HOTI.

%================================================

%==================================

\section*{Methods}

\textbf{Sample Preparation}: Highly crystalline films of Bi and  Bi$_{0.92}$Sb$_{0.08}$ were grown on n-doped silicon (111) wafers using a custom ultrahigh vacuum molecular beam epitaxy system. Before deposition, Si substrates were cleaned by ultrasonication in acetone and isopropanol. The substrates were then transferred to the MBE system (pressure $<2\times10^{-9}$ Torr) and degassed at $520^{\circ}C$ for 10-12 hrs. After degassing, the substrate was cleaned by flash annealing where they were repeatedly heated from $650^{\circ}C$ to $900^{\circ}C$ in 80 s. Each cycle helps to remove oxide layers or impurities from the silicon surface. The  surface quality was confirmed by the observation of clear $7\times7$ reconstructed RHEED patterns. The substrate was rapidly cooled from $650^{\circ}C$ to room temperature at the rate of $1-2^{\circ}C/s$ after 16 cycles of flash annealing. For film growth, bismuth (purity $\sim 99.9999 \% $) and antimony (purity $\sim 99.9999 \% $) sources were evaporated from standard Knudcell cells at flux rates 0.0769 $\mathring{A}/s$ and 0.0071 $\mathring{A}/s$ respectively as measured by a Quartz Crystal monitor. The substrates were held at room temperature during growth. 
\\

\noindent\textbf{STM measurements}: Samples were transferred from MBE chamber to a vacuum suitcase at pressure $<3\times10^{-10}$ and subsequently transferred to a Unisoku UHV-STM without exposure to air. All the measurements were done at 4.3 K. Normal tungsten tips were prepared by etching method and used after cleaning by electron beam heating. All the topographic images were taken in constant-current mode and spectroscopic signals were recorded by standard lock-in amplifier with ac modulation 3-3.5 mV at 983 Hz frequency.

\vspace{24pt}

% \newpage

%\bibliographystyle{naturemag}
%%{\footnotesize
    %\bibliography{ref}

\begin{thebibliography}{10}
\expandafter\ifx\csname url\endcsname\relax
  \def\url#1{\texttt{#1}}\fi
\expandafter\ifx\csname urlprefix\endcsname\relax\def\urlprefix{URL }\fi
\providecommand{\bibinfo}[2]{#2}
\providecommand{\eprint}[2][]{\url{#2}}

\bibitem{schindler2018higher}
\bibinfo{author}{Schindler, F.} \emph{et~al.}
\newblock \bibinfo{title}{Higher-order topology in bismuth}.
\newblock \emph{\bibinfo{journal}{Nature Physics}}
  \textbf{\bibinfo{volume}{14}}, \bibinfo{pages}{918} (\bibinfo{year}{2018}).

\bibitem{song2017d}
\bibinfo{author}{Song, Z.}, \bibinfo{author}{Fang, Z.} \&
  \bibinfo{author}{Fang, C.}
\newblock \bibinfo{title}{(d-2)-dimensional edge states of rotation symmetry
  protected topological states}.
\newblock \emph{\bibinfo{journal}{Physical review letters}}
  \textbf{\bibinfo{volume}{119}}, \bibinfo{pages}{246402}
  (\bibinfo{year}{2017}).

\bibitem{benalcazar2018prb}
\bibinfo{author}{Benalcazar, W.~A.}, \bibinfo{author}{Bernevig, B.~A.} \&
  \bibinfo{author}{Hughes, T.~L.}
\newblock \bibinfo{title}{Electric multipole moments, topological multipole
  moment pumping, and chiral hinge states in crystalline insulators}.
\newblock \emph{\bibinfo{journal}{Physical Review B}}
  \textbf{\bibinfo{volume}{96}}, \bibinfo{pages}{245115}
  (\bibinfo{year}{2017}).

\bibitem{Schindlereaat0346}
\bibinfo{author}{Schindler, F.} \emph{et~al.}
\newblock \bibinfo{title}{Higher-order topological insulators}.
\newblock \emph{\bibinfo{journal}{Science Advances}}
  \textbf{\bibinfo{volume}{4}} (\bibinfo{year}{2018}).

\bibitem{Olsen2020gaplss}
\bibinfo{author}{Olsen, T.}, \bibinfo{author}{Rauch, T. c.~v.},
  \bibinfo{author}{Vanderbilt, D.} \& \bibinfo{author}{Souza, I.}
\newblock \bibinfo{title}{Gapless hinge states from adiabatic pumping of axion
  coupling}.
\newblock \emph{\bibinfo{journal}{Phys. Rev. B}}
  \textbf{\bibinfo{volume}{102}}, \bibinfo{pages}{035166}
  (\bibinfo{year}{2020}).
\newblock \urlprefix\url{https://link.aps.org/doi/10.1103/PhysRevB.102.035166}.

\bibitem{zhu2020quantized}
\bibinfo{author}{Zhu, P.}, \bibinfo{author}{Hughes, T.~L.} \&
  \bibinfo{author}{Alexandradinata, A.}
\newblock \bibinfo{title}{Quantized surface magnetism and higher-order
  topology: Application to the hopf insulator}.
\newblock \emph{\bibinfo{journal}{Phys. Rev. B}}
  \textbf{\bibinfo{volume}{103}}, \bibinfo{pages}{014417}
  (\bibinfo{year}{2021}).
\newblock \urlprefix\url{https://link.aps.org/doi/10.1103/PhysRevB.103.014417}.

\bibitem{serra2018observation}
\bibinfo{author}{Serra-Garcia, M.} \emph{et~al.}
\newblock \bibinfo{title}{Observation of a phononic quadrupole topological
  insulator}.
\newblock \emph{\bibinfo{journal}{Nature}} \textbf{\bibinfo{volume}{555}},
  \bibinfo{pages}{342--345} (\bibinfo{year}{2018}).

\bibitem{peterson2018quantized}
\bibinfo{author}{Peterson, C.~W.}, \bibinfo{author}{Benalcazar, W.~A.},
  \bibinfo{author}{Hughes, T.~L.} \& \bibinfo{author}{Bahl, G.}
\newblock \bibinfo{title}{A quantized microwave quadrupole insulator with
  topologically protected corner states}.
\newblock \emph{\bibinfo{journal}{Nature}} \textbf{\bibinfo{volume}{555}},
  \bibinfo{pages}{346--350} (\bibinfo{year}{2018}).

\bibitem{noh2018topological}
\bibinfo{author}{Noh, J.} \emph{et~al.}
\newblock \bibinfo{title}{Topological protection of photonic mid-gap defect
  modes}.
\newblock \emph{\bibinfo{journal}{Nature Photonics}}
  \textbf{\bibinfo{volume}{12}}, \bibinfo{pages}{408--415}
  (\bibinfo{year}{2018}).

\bibitem{imhof2018topolectrical}
\bibinfo{author}{Imhof, S.} \emph{et~al.}
\newblock \bibinfo{title}{Topolectrical-circuit realization of topological
  corner modes}.
\newblock \emph{\bibinfo{journal}{Nature Physics}}
  \textbf{\bibinfo{volume}{14}}, \bibinfo{pages}{925--929}
  (\bibinfo{year}{2018}).

\bibitem{choi2020evidence}
\bibinfo{author}{Choi, Y.-B.} \emph{et~al.}
\newblock \bibinfo{title}{Evidence of higher-order topology in multilayer wte 2
  from josephson coupling through anisotropic hinge states}.
\newblock \emph{\bibinfo{journal}{Nature Materials}}
  \textbf{\bibinfo{volume}{19}}, \bibinfo{pages}{974--979}
  (\bibinfo{year}{2020}).

\bibitem{lee2020two}
\bibinfo{author}{Lee, E.}, \bibinfo{author}{Kim, R.}, \bibinfo{author}{Ahn, J.}
  \& \bibinfo{author}{Yang, B.-J.}
\newblock \bibinfo{title}{Two-dimensional higher-order topology in monolayer
  graphdiyne}.
\newblock \emph{\bibinfo{journal}{npj Quantum Materials}}
  \textbf{\bibinfo{volume}{5}}, \bibinfo{pages}{1--7} (\bibinfo{year}{2020}).

\bibitem{zhang2020mobius}
\bibinfo{author}{Zhang, R.-X.}, \bibinfo{author}{Wu, F.} \&
  \bibinfo{author}{Sarma, S.~D.}
\newblock \bibinfo{title}{M{\"o}bius insulator and higher-order topology in
  $\mathrm{MnBi}_{2 n} \mathrm{Te}_{3 n+1}$}.
\newblock \emph{\bibinfo{journal}{Physical Review Letters}}
  \textbf{\bibinfo{volume}{124}}, \bibinfo{pages}{136407}
  (\bibinfo{year}{2020}).

\bibitem{wang2019higher}
\bibinfo{author}{Wang, Z.}, \bibinfo{author}{Wieder, B.~J.},
  \bibinfo{author}{Li, J.}, \bibinfo{author}{Yan, B.} \&
  \bibinfo{author}{Bernevig, B.~A.}
\newblock \bibinfo{title}{Higher-order topology, monopole nodal lines, and the
  origin of large fermi arcs in transition metal dichalcogenides $\boldsymbol{X T e}_{2}(\boldsymbol{X}=\mathbf{M o}, \mathbf{W})$}.
\newblock \emph{\bibinfo{journal}{Physical review letters}}
  \textbf{\bibinfo{volume}{123}}, \bibinfo{pages}{186401}
  (\bibinfo{year}{2019}).

\bibitem{zhang2019higher}
\bibinfo{author}{Zhang, R.-X.}, \bibinfo{author}{Cole, W.~S.},
  \bibinfo{author}{Wu, X.} \& \bibinfo{author}{Sarma, S.~D.}
\newblock \bibinfo{title}{Higher-order topology and nodal topological
  superconductivity in Fe(Se,Te) heterostructures}.
\newblock \emph{\bibinfo{journal}{Physical review letters}}
  \textbf{\bibinfo{volume}{123}}, \bibinfo{pages}{167001}
  (\bibinfo{year}{2019}).

\bibitem{park2019higher}
\bibinfo{author}{Park, M.~J.}, \bibinfo{author}{Kim, Y.}, \bibinfo{author}{Cho,
  G.~Y.} \& \bibinfo{author}{Lee, S.}
\newblock \bibinfo{title}{Higher-order topological insulator in twisted bilayer
  graphene}.
\newblock \emph{\bibinfo{journal}{Physical Review Letters}}
  \textbf{\bibinfo{volume}{123}}, \bibinfo{pages}{216803}
  (\bibinfo{year}{2019}).

\bibitem{drozdov2014one}
\bibinfo{author}{Drozdov, I.~K.} \emph{et~al.}
\newblock \bibinfo{title}{One-dimensional topological edge states of bismuth
  bilayers}.
\newblock \emph{\bibinfo{journal}{Nature Physics}}
  \textbf{\bibinfo{volume}{10}}, \bibinfo{pages}{664--669}
  (\bibinfo{year}{2014}).

\bibitem{hsu2019topology}
\bibinfo{author}{Hsu, C.-H.} \emph{et~al.}
\newblock \bibinfo{title}{Topology on a new facet of bismuth}.
\newblock \emph{\bibinfo{journal}{Proceedings of the National Academy of
  Sciences}} \textbf{\bibinfo{volume}{116}}, \bibinfo{pages}{13255--13259}
  (\bibinfo{year}{2019}).

\bibitem{takayama2015one}
\bibinfo{author}{Takayama, A.}, \bibinfo{author}{Sato, T.},
  \bibinfo{author}{Souma, S.}, \bibinfo{author}{Oguchi, T.} \&
  \bibinfo{author}{Takahashi, T.}
\newblock \bibinfo{title}{One-dimensional edge states with giant spin splitting
  in a bismuth thin film}.
\newblock \emph{\bibinfo{journal}{Physical review letters}}
  \textbf{\bibinfo{volume}{114}}, \bibinfo{pages}{066402}
  (\bibinfo{year}{2015}).

\bibitem{kawakami2015one}
\bibinfo{author}{Kawakami, N.}, \bibinfo{author}{Lin, C.-L.},
  \bibinfo{author}{Kawai, M.}, \bibinfo{author}{Arafune, R.} \&
  \bibinfo{author}{Takagi, N.}
\newblock \bibinfo{title}{One-dimensional edge state of bi thin film grown on
  si (111)}.
\newblock \emph{\bibinfo{journal}{Applied Physics Letters}}
  \textbf{\bibinfo{volume}{107}}, \bibinfo{pages}{031602}
  (\bibinfo{year}{2015}).

\bibitem{nagao2004nanofilm}
\bibinfo{author}{Nagao, T.} \emph{et~al.}
\newblock \bibinfo{title}{Nanofilm allotrope and phase transformation of
  ultrathin bi film on $\mathrm{Si}(111)-7 \times 7$}.
\newblock \emph{\bibinfo{journal}{Physical review letters}}
  \textbf{\bibinfo{volume}{93}}, \bibinfo{pages}{105501}
  (\bibinfo{year}{2004}).

\bibitem{PhysRevB.91.075429}
\bibinfo{author}{Kokubo, I.}, \bibinfo{author}{Yoshiike, Y.},
  \bibinfo{author}{Nakatsuji, K.} \& \bibinfo{author}{Hirayama, H.}
\newblock \bibinfo{title}{Ultrathin bi(110) films on
  $\mathrm{Si}(111)\sqrt{3}\ifmmode\times\else\times\fi{}\sqrt{3}\text{\ensuremath{-}}\mathrm{B}$
  substrates}.
\newblock \emph{\bibinfo{journal}{Phys. Rev. B}} \textbf{\bibinfo{volume}{91}},
  \bibinfo{pages}{075429} (\bibinfo{year}{2015}).
\newblock \urlprefix\url{https://link.aps.org/doi/10.1103/PhysRevB.91.075429}.

\bibitem{2020APExp..13h5506N}
\bibinfo{author}{{Nagase}, K.}, \bibinfo{author}{{Ushioda}, R.},
  \bibinfo{author}{{Nakatsuji}, K.}, \bibinfo{author}{{Shirasawa}, T.} \&
  \bibinfo{author}{{Hirayama}, H.}
\newblock \bibinfo{title}{{Growth of extremely flat Bi(110) films on a
  $\mathrm{Si}(111)\sqrt{3}\ifmmode\times\else\times\fi{}\sqrt{3}\text{\ensuremath{-}}\mathrm{B}$
  substrate}}.
\newblock \emph{\bibinfo{journal}{Applied Physics Express}}
  \textbf{\bibinfo{volume}{13}}, \bibinfo{pages}{085506}
  (\bibinfo{year}{2020}).

\bibitem{takayama2014rashba}
\bibinfo{author}{Takayama, A.}, \bibinfo{author}{Sato, T.},
  \bibinfo{author}{Souma, S.} \& \bibinfo{author}{Takahashi, T.}
\newblock \bibinfo{title}{Rashba effect in antimony and bismuth studied by
  spin-resolved arpes}.
\newblock \emph{\bibinfo{journal}{New Journal of Physics}}
  \textbf{\bibinfo{volume}{16}}, \bibinfo{pages}{055004}
  (\bibinfo{year}{2014}).

\bibitem{tanaka2012experimental}
\bibinfo{author}{Tanaka, Y.} \emph{et~al.}
\newblock \bibinfo{title}{Experimental realization of a topological crystalline
  insulator in snte}.
\newblock \emph{\bibinfo{journal}{Nature Physics}}
  \textbf{\bibinfo{volume}{8}}, \bibinfo{pages}{800--803}
  (\bibinfo{year}{2012}).

\bibitem{lenoir1996bi}
\bibinfo{author}{Lenoir, B.} \emph{et~al.}
\newblock \bibinfo{title}{Bi-sb alloys: an update}.
\newblock In \emph{\bibinfo{booktitle}{Fifteenth International Conference on
  Thermoelectrics. Proceedings ICT'96}}, \bibinfo{pages}{1--13}
  (\bibinfo{organization}{IEEE}, \bibinfo{year}{1996}).

\bibitem{fu2007topological}
\bibinfo{author}{Fu, L.}, \bibinfo{author}{Kane, C.~L.} \&
  \bibinfo{author}{Mele, E.~J.}
\newblock \bibinfo{title}{Topological insulators in three dimensions}.
\newblock \emph{\bibinfo{journal}{Phys. Rev. Lett.}}
  \textbf{\bibinfo{volume}{98}}, \bibinfo{pages}{106803}
  (\bibinfo{year}{2007}).
\newblock
  \urlprefix\url{https://link.aps.org/doi/10.1103/PhysRevLett.98.106803}.

\bibitem{teo2008surface}
\bibinfo{author}{Teo, J.~C.}, \bibinfo{author}{Fu, L.} \&
  \bibinfo{author}{Kane, C.}
\newblock \bibinfo{title}{Surface states and topological invariants in
  three-dimensional topological insulators: Application to $\mathbf{B i}_{\mathbf{1}-\boldsymbol{x}} \mathbf{S} \mathbf{b}_{\boldsymbol{x}}$}.
\newblock \emph{\bibinfo{journal}{Physical Review B}}
  \textbf{\bibinfo{volume}{78}}, \bibinfo{pages}{045426}
  (\bibinfo{year}{2008}).

\end{thebibliography}
%%}
%============================================
\section*{Acknowledgments} STM studies and theoretical work at the University of Illinois, Urbana-Champaign were primarily supported by the Materials Research Science and Engineering Center DMR-1720633. VM acknowledges support from the Gordon and Betty Moore Foundation’s EPiQS Initiative, Grant GBMF9465 for MBE growth and characterization. The authors thank G. Chen, Z. Cai, Y. Maximenko and C. Steiner for valuable discussions. Thin film characterization was carried out in part in the Materials Research Laboratory Central Research Facilities at the University of Illinois. 

\section*{Author contributions} T.L.H and V.M. conceived the idea, designed the experiments and developed the analysis tools. L.A. grew and characterized all the materials and carried out the STM/STS experiments. P.Z. carried out the theoretical analysis. V.M. and T.L.H supervised all aspects of the project. All authors jointly wrote the paper.

%============================================

\clearpage

\clearpage

\clearpage

\newcommand{\beginsupplement}{%
        \setcounter{table}{0}
        \renewcommand{\thetable}{S\arabic{table}}%
        \setcounter{figure}{0}
        \renewcommand{\thefigure}{S\arabic{figure}}%
        \setcounter{equation}{0}
        \renewcommand{\theequation}{S\arabic{equation}}%\
        \setcounter{section}{0}
        \renewcommand{\thesection}{S\arabic{section}}%
}

\end{document}

\title{ Supplementary information for 
\\
Evidence for Higher order topology in B\lowercase{i} and B\lowercase{i}$_{0.92}$S\lowercase{b}$_{0.08}$}
\author{Leena Aggarwal}
\affiliation{Department of Physics and Materials Research Laboratory, University of Illinois
Urbana-Champaign, Urbana, Illinois 61801, USA}
\author{Penghao Zhu}
\affiliation{Department of Physics and Institute for Condensed Matter Theory,
University of Illinois at Urbana-Champaign, Illinois 61801, USA}
\author{Taylor L. Hughes}
\affiliation{Department of Physics and Institute for Condensed Matter Theory,
University of Illinois at Urbana-Champaign, Illinois 61801, USA}
\author{Vidya Madhavan}\altaffiliation{To whom correspondence should be addressed: vm1@illinois.edu}
\affiliation{Department of Physics and Materials Research Laboratory, University of Illinois
Urbana-Champaign, Urbana, Illinois 61801, USA}

%\date{\today}
\date{\today}
\maketitle

\onecolumngrid

\section*{Supplementary note 1: 1D hinge modes at type-A edges in Bi(111) film}

\begin{figure}[h]
\centering
\includegraphics[width=0.75\columnwidth]{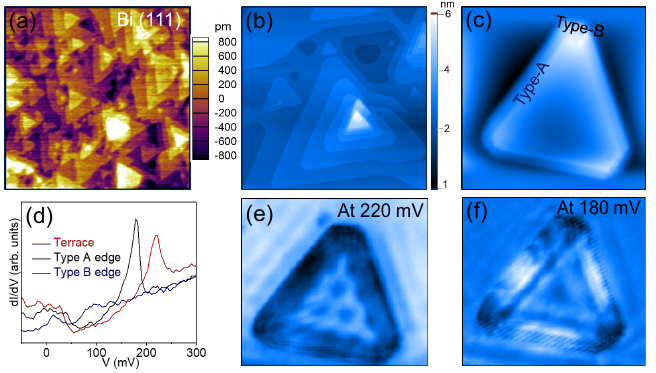}
\caption{\textbf{STM studies on Bi(111) films grown on  silicon (111)substrates:} (a) A large area (700 nm $\times$700 nm) atomic force microscopy (AFM) image  showing triangular islands which confirms the (111) orientation of the bismuth film with the color bar showing the relative height.(b) 300 nm $\times$300 nm STM image of the film with the color bar showing the relative height. (c) Topography of a 28 nm $\times$28 nm triangular (111) facet with two kind of edges, marked as Type-A and Type-B. (d) dI/dV spectra on the two types of edges and near the center of the island.  The hinge mode density of states peak can be seen at the type-A edge. (e,f) STM dI/dV maps at 220 mV and 180 mV. The 180mV map clearly shows the localization of hinge modes at the three type-A edges.}
\label{fig:hinge_Bi}
\end{figure}

\section*{Supplementary note 2: Theoretical explanation for helical edge modes on the (110) facet}
\subsection*{Orientation of surface facets}

\label{app: orientation}
\begin{figure}[h]
\centering
\includegraphics[width=0.8\columnwidth]{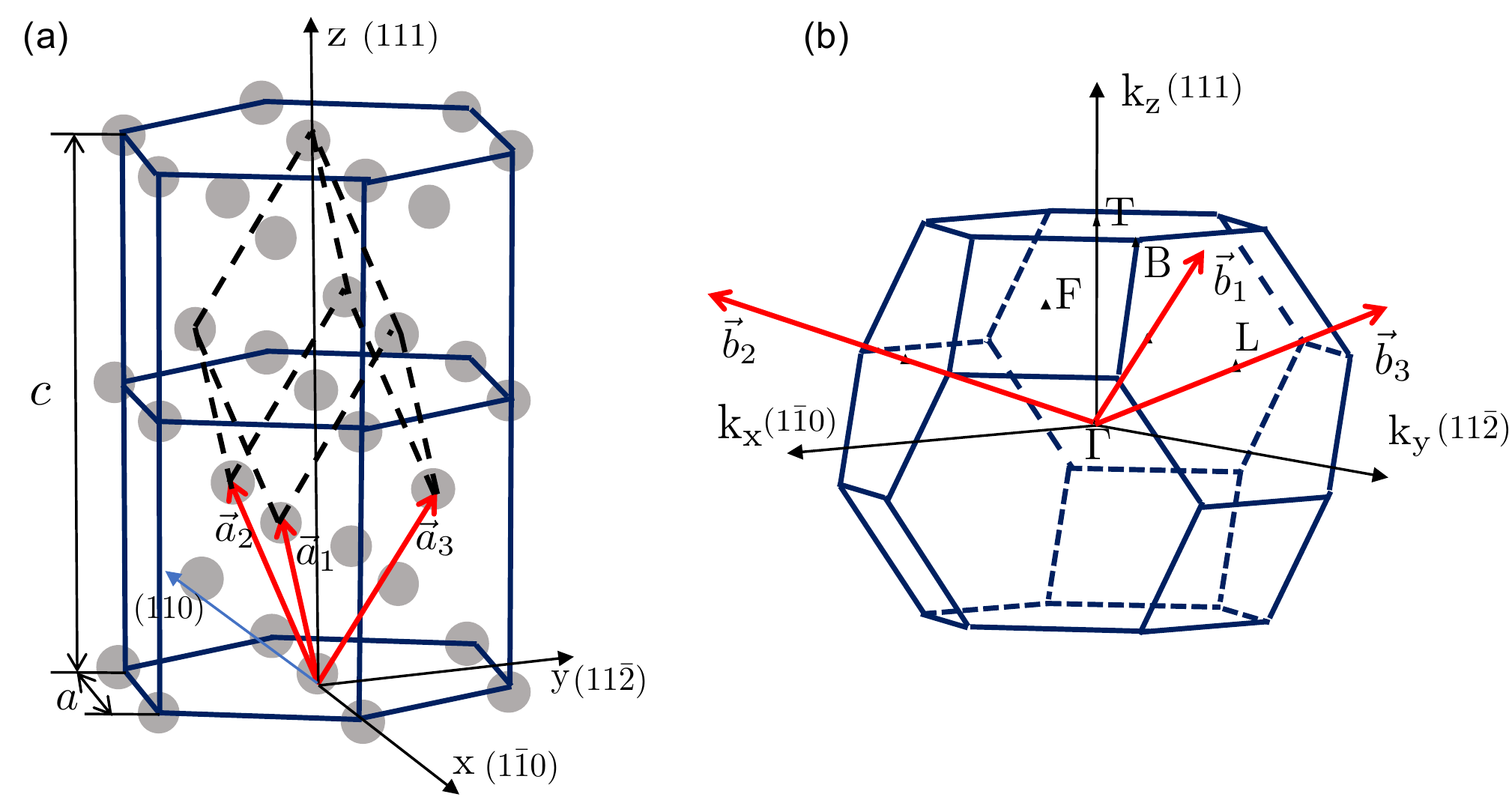}
\caption{(a) Crystal structure of Bi and/or BiSb alloys with primitive cell, indicating rhombohedral vectors  ($\vec{a}_{1}$,$\vec{a}_{2}$,$\vec{a}_{3}$) and $(11\bar{2})$, $(111)$, $(1\bar{1}0)$, $(110)$ planes. (b) First Brillouin zone of Bi and/or BiSb alloys band structure.}
\label{fig: lattice_BZ}
\end{figure}

As shown in Fig. \ref{fig: lattice_BZ} (a), in the Cartesian coordinate system we choose, the lattice vectors can be expressed as
\begin{equation}
\label{eq: latticevec}
\vec{a}_{1}=(-\frac{1}{2} a,-\frac{\sqrt{3}}{6} a, \frac{1}{3} c),\quad \vec{a}_{2}=(\frac{1}{2} a,-\frac{\sqrt{3}}{6} a, \frac{1}{3} c),\quad \vec{a}_{3}=(0, \frac{\sqrt{3}}{3} a, \frac{1}{3} c),
\end{equation}
where $a$ and $c$ are lattice constants as illustrated in Fig. \ref{fig: lattice_BZ} (a). As shown in Fig. \ref{fig: lattice_BZ} (b), the three corresponding reciprocal-lattice vectors defined by $\bm{b}_{i} \cdot \bm{a}_{j}=2 \pi \delta_{i j}$ with $i,j=1,2,3$, are given by 
\begin{equation}
\label{eq: rlatticevec}
\vec{b}_{1}=g(-1,-\frac{\sqrt{3}}{3}, \frac{a}{c}),
\quad \vec{b}_{2}=g(1,-\frac{\sqrt{3} }{3} , \frac{a}{c}),
\quad \vec{b}_{3}=g(0,\frac{2 \sqrt{3}}{3} , \frac{a}{c}),
\end{equation}
where $g=2\pi/a$. From Eq. \eqref{eq: rlatticevec}, we can derive that
\begin{equation}
\label{eq: vector}
\begin{array}{l}
(111)=\vec{b}_{1}+\vec{b}_{2}+\vec{b}_{3}=g(0,0 , 3a/c),
\\
(1\bar{1}0)=\vec{b}_{1}-\vec{b}_{2}=g(-2,0 , 0),
\\
(11\bar{2})=\vec{b}_{1}+\vec{b}_{2}-2\vec{b}_{3}=g(0,-2\sqrt{3},0),
\\
(110)=\vec{b}_{1}+\vec{b}_{2}=g(0,-2\sqrt{3}/3 , 2a/c),
\end{array}
\end{equation}
from which we see that $(1\bar{1}0)$, $(11\bar{2})$, and $(111)$ are the Cartesian $x$, $y$, and $z$ directions (up to a sign), respectively. We can also see that $(110)$, $(1\bar{1}0)$ and $\vec{a}_{3}$ are orthogonal to each other.

\subsection*{Possible configurations of hinge modes under different symmetries}
\begin{figure}[h]
\centering
\includegraphics[width=0.9\columnwidth]{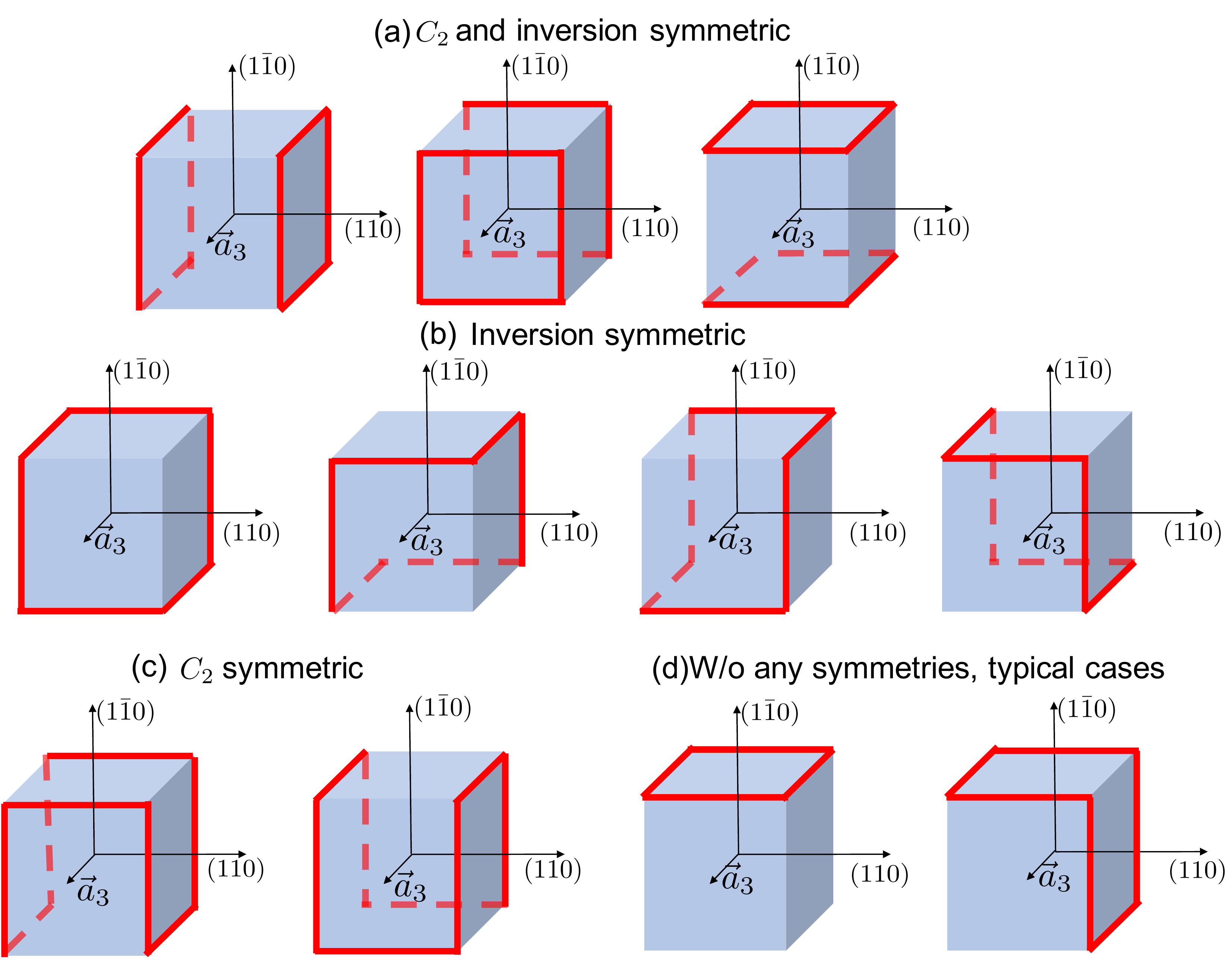}
\caption{All possible configurations of helical hinge modes on the (110) surface of $\rm{Bi}$ and/or $\rm{Bi_{0.92}Sb_{0.08}}$ when (a) both $\hat{C}_{2}$ and inversion symmetries are preserved, (b) inversion is preserved but $\hat{C}_{2}$ is broken, and (c) $\hat{C}_{2}$ is preserved but inversion is broken. (d) shows two typical configurations when both $\hat{C}_{2}$ and inversion symmetries are broken. The red-color hinges are hinges where the surface mass vanish, $i.e.$, where there are gapless hinge modes. Note that $\hat{C}_{2}$ here refers to the two-fold rotation around $(1\bar{1}0)$ axis.}
\label{fig:enumerate}
\end{figure}

As discussed in the main text, the exotic three-sided hinge modes configuration observed on the rectangular island of a (110) surface facet is ascribed to higher order topology protected by the $\hat{C}_{2}$ symmetry around the  $(1\bar{1}0)$-axis, and the time-reversal symmetry. We claim that it is necessary to break inversion symmetry to observe this hinge modes configuration. To support this claim, we enumerate all possible configurations of edge states when both $\hat{C}_{2}$ and inversion symmetries are preserved, only $\hat{C}_{2}$ is preserved, and only inversion is preserved. We also show two typical configurations which break both inversion and $\hat{C}_{2}$ rotation symmetries, one of which also gives the three-sided configurations. The enumeration shown in Fig. \ref{fig:enumerate} clearly show that the three-sided edge modes configurations can only happen when the inversion symmetry is broken, irrespective of whether the $\hat{C}_{2}$ rotation symmetry is broken or not.

\clearpage

\section*{Supplementary note 3: Raw Data on Bi(110) and Bi$_{0.92}$Sb$_{0.08}$(110) Films}
\begin{figure}[h]
\centering
\includegraphics[width=0.85\columnwidth]{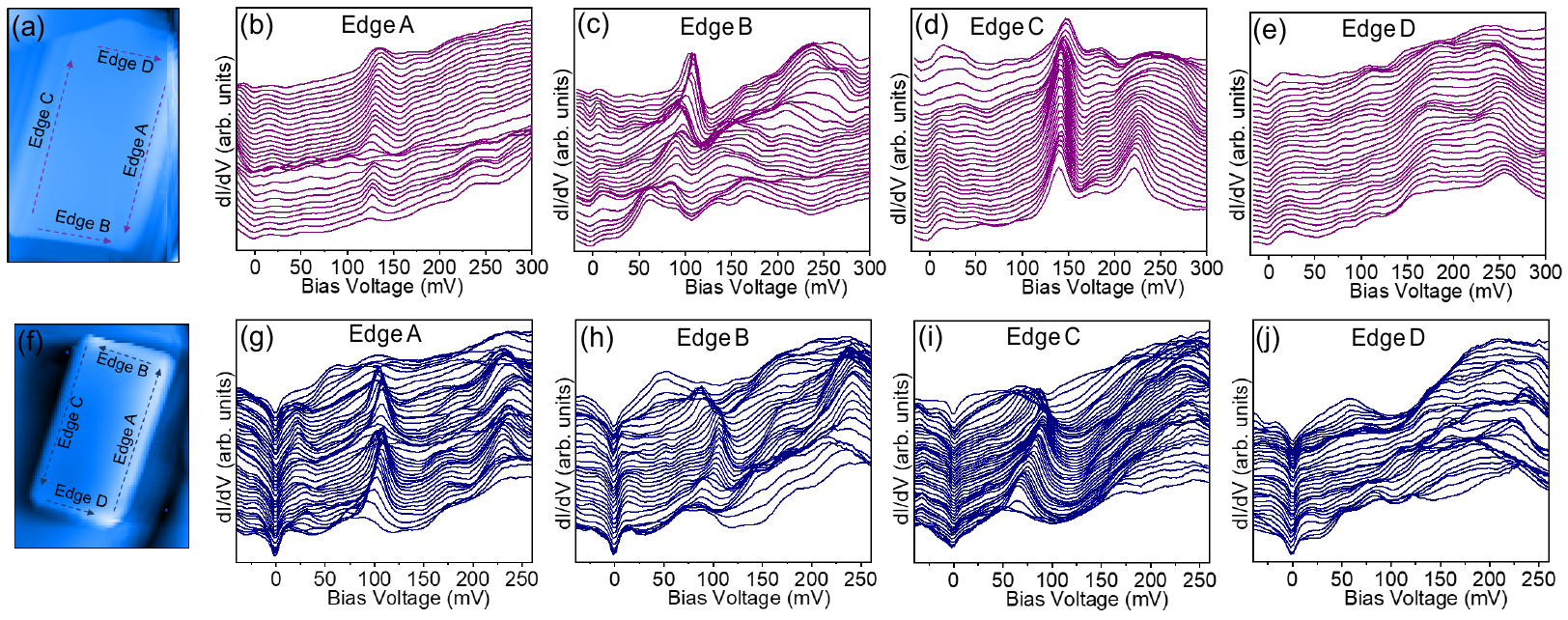}
\caption{(a) Topographic image of the of same Bi(110) rectangular island shown in Figure 3 (a) in the main manuscript. (b,c,d,e)  Raw spectroscopic data (without slope subtraction) (at ac modulation 3.5 mV and current 60 pA) along the four edges marked Edge A , Edge B, Edge C and Edge D (indicated by purple arrows in Figure S4(a)). (f) Topographic image of the same Bi$_{0.92}$Sb$_{0.08}$(110) rectangular island as shown in Figure 4 (e) in the main manuscript. (g,h,i,j) Raw spectroscopic data (without slope subtraction) (at ac modulation 3 mV and current 120 pA) along the four edges marked Edge A, Edge B, Edge C and Edge D ( shown as blue arrows along the edges in Figure S4(f)).}
\end{figure}

\section*{Supplementary note 4: Rutherford Backscattering Spectrometery (RBS) to confirm Sb composition in alloy}

\begin{figure}[h]
\centering
\includegraphics[width=0.50\columnwidth]{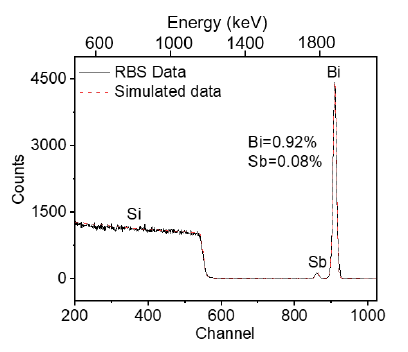}
\caption{RBS data (in solid black color) on BiSb alloy to confirm the composition of Sb with nicely fitted simulated data (in dashed red color line). Fitting confirms Sb composition 0.08$\%$. }
\label{fig:rbs}
\end{figure}

\newpage

\section*{Supplementary note 5: Additional Raw Data on Bi(110) and Bi$_{0.92}$Sb$_{0.08}$(110) Films}

\begin{figure}[h]
\centering
\includegraphics[width=0.85\columnwidth]{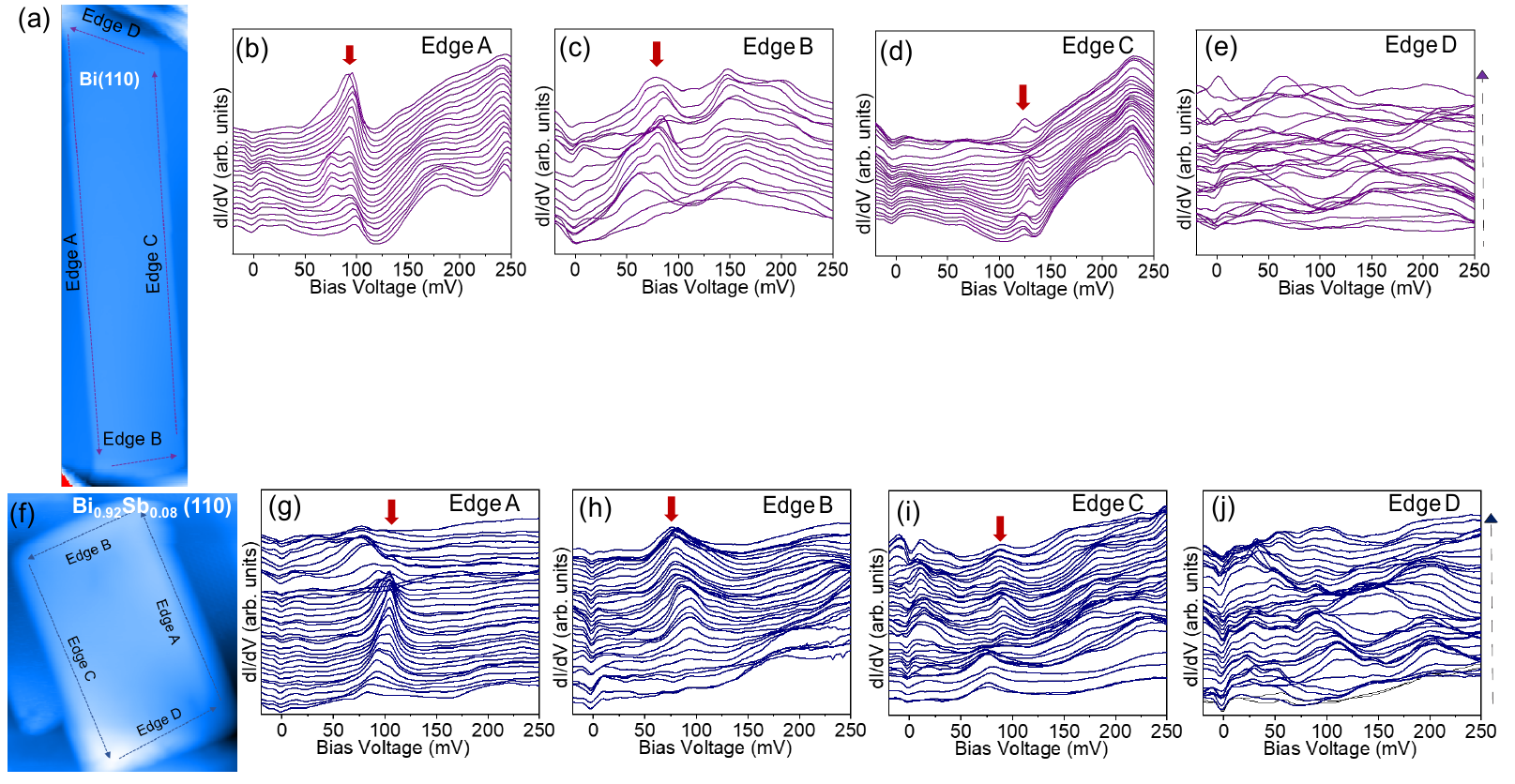}
\caption{(a)Topographic image of the of same Bi(110) rectangular island. (b,c,d,e)  Raw spectroscopic data (without slope subtraction) (at ac modulation 3.5 mV and current 80 pA) along the four edges marked Edge A , Edge B, Edge C and Edge D (indicated by purple arrows in Figure S6(a)). (f) Topographic image of the same Bi$_{0.92}$Sb$_{0.08}$(110) rectangular island. (g,h,i,j) Raw spectroscopic data (without slope subtraction) (at ac modulation 2.5 mV and current 120 pA) along the four edges marked Edge A, Edge B, Edge C and Edge D ( shown as blue arrows along the edges in Figure S6(f)). The red block arrows indicate the sharp peaks corresponding to the edge modes that are only seen on three of the four edges.}

\end{figure}

\newpage
\section*{Supplementary note 6: Schematic of Bi(110) lattice structure} 

The schematic of Bi(110) lattice structure, as shown in Figure S7(a), aligned correctly with the long edge  and short edge of (110) island. The atoms along the two long edges  of the (110) island bond differently to the layer below. However, in our data, both long edges of the (110) islands always show edge states. Put another way, the two short edge sides are identical from a bonding perspective. But, for a given island, one short side shows hinge states while the other does not. Since the appearance hinge modes at three out of four edges has no connection with the geometry of the bonds at the edges, we can rule out bonding effects as an explanation for the pattern of our edge modes. Our data are instead well explained by the Higher order topological model.

\begin{figure}[h]
\centering
\includegraphics[width=0.65\columnwidth]{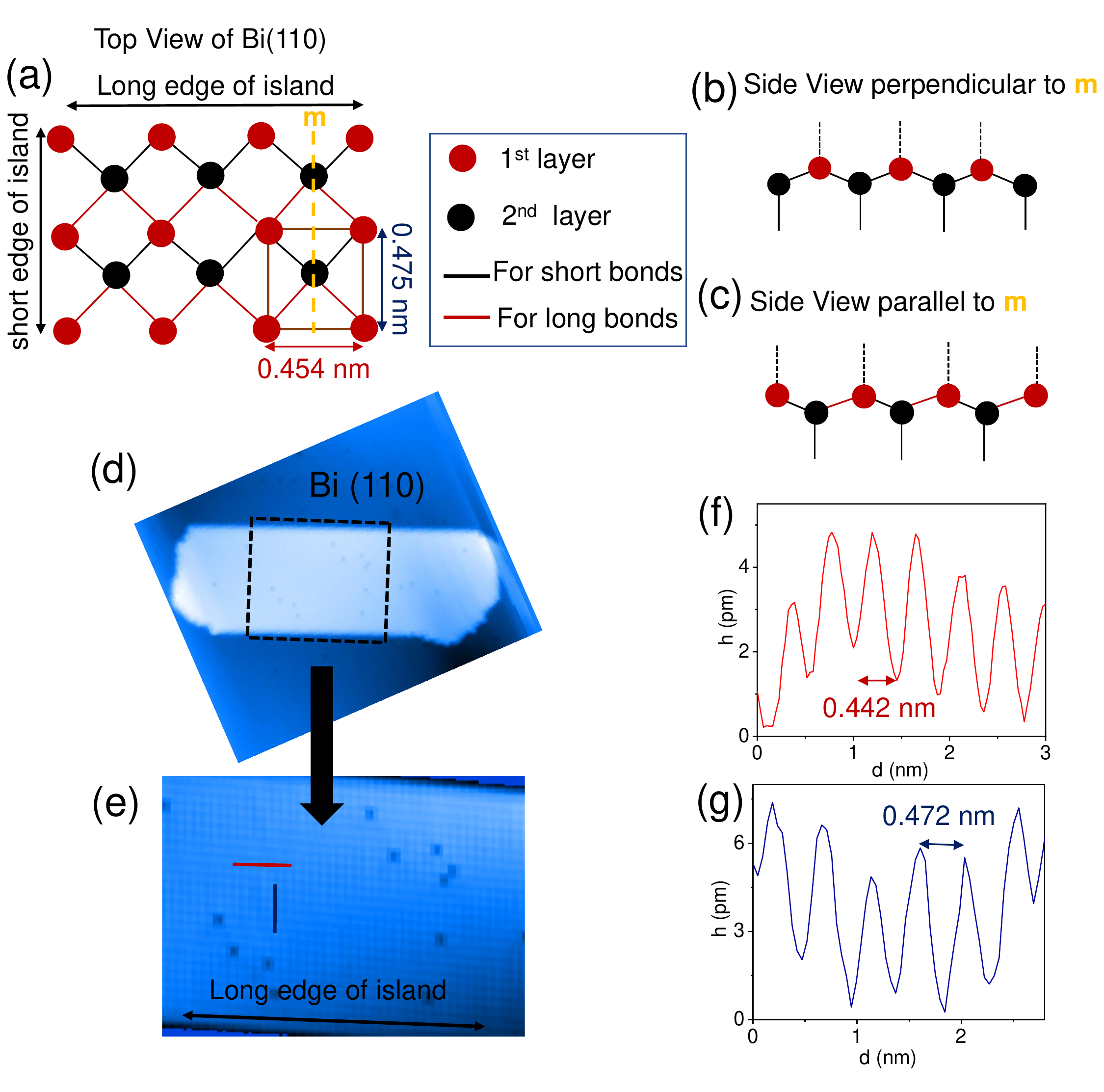}
\caption{(a) Schematic of the top view of first two layers of  Bi(110) lattice structure, indicating two long edges of island bonded differently. Red atoms shows the first layer of two-dimensional rectangular lattice with lattice constants. The red lines and black lines between atoms represent long and short bonds respectively. The mirror plane (`m') of the structure are also shown as dashed yellow line.The lattice constants along long edge of island is 0.454 nm and short edge of island is 0.472 nm.  (b, c) Side view of the lattice structure perpendicular and parallel to mirror plane (`m') respectively.(d) STM image of Bi(110) island shows the long and short edge sides of island same as in (a). (e) Zoomed image of Bi(110) island shown in (d), showing pseudo-cubic arrangements of atoms. (f,g) Lattice constants along the long edge side (corresponding red line in (e)) of island and short edge side (corresponding blue line in (e)) of the island respectively.}
\end{figure}